\documentclass{article}
\usepackage{amsfonts,amssymb, amsmath}
\usepackage[all]{xy}
\usepackage[english]{babel}

\def\nn{\nonumber }
\def\bq{ \begin{equation} }
\def\eq{ \end{equation} }
\def\ben{ \begin{eqnarray} }
\def\en{ \end{eqnarray} }

\newtheorem{prop}{Proposition}

\newtheorem{re}{Remark}
\newenvironment{rem}{\begin{re} \rm }{\end{re}}
\begin{document}


\title{Simultaneous separation for the Neumann and Chaplygin systems.}

\author{A. V. Tsiganov\\
\it\small
St.Petersburg State University, St.Petersburg, Russia\\
\it\small e--mail: andrey.tsiganov@gmail.com}

\date{}
\maketitle

\begin{abstract}
The Neumann and Chaplygin systems on the sphere are simultaneously separable in variables obtained from the standard
elliptic coordinates by the proper B\"{a}cklund transformation. We also prove that after similar B\"{a}cklund transformations other curvilinear coordinates on the sphere and on the plane become variables of separations for the system with quartic potential, for the H\'{e}non-Heiles system and for the Kowalevski top.  It allows us to say about some analog of the hetero B\"{a}cklund transformations relating different Hamilton-Jacobi equations.

 \end{abstract}

\section{Introduction}
\setcounter{equation}{0}
The classical Neumann system describes the dynamics of a particle constrained to move on the sphere under the influence of a quadratic potential
\[ V (x) =a_1x_1^2 + a_2x_2^2 + a_3x_3^2,\]
where ($x_1, x_2, x_3$)  are Cartesian coordinates in three-dimensional Euclidean space \cite{neu59}.
Generally the Neumann model represents a harmonic oscillator restricted on  $(n-1)$-dimensional sphere in $n$ dimensional Euclidean space. A fairly large body of literature has been devoted to this system, which is studied well in the framework of the dynamical system \cite{mos80-1,mos80-2}, of the symplectic geometry \cite{ves01,gs84}, of the algebraic geometry \cite{mum84}, of the representation of the infinite Lie algebra \cite{harn88}, of the quantum mechanics \cite{bab92,bell05,gur95}, etc.

In 1859 Carl Neumann showed that the equations of motion could be solved using the Jacobi theory of separation of variables \cite{neu59}. More than one century later, this separability result was generalized to the arbitrary $n$-dimensional case by Moser \cite{mos80-1}. The starting point to solve the problem was the ingenious introduction of a special set of coordinates called spheroconical or elliptical coordinates on the sphere. Later on it was shown that these coordinates could be framed within the formalism of Lax pairs, $r$-matrices (see, e.g., \cite{harn93,bab92}) and also within the bi-Hamiltonian formalism \cite{ped04}.

 In this article we calculate other variables of separation for the Neumann system using the notions of bi-Hamiltonian geometry, generalizing and refining the approach described in \cite{ts11r,ts10k,ts11,ts11s}. We will start with the given integrals of motion $H_{1,2}$ and look at the bi-involution condition
\[
\{H_1,H_2\}=\{H_1,H_2\}'=0
\]
as an equation to determine the second Poisson bracket $\{.,.\}'$ and, hence, the separation coordinates. We shall see that it is indeed possible and, actually easy, to solve such an equation by means of a couple of natural
Ans\"atze, thus arriving to induce the separation coordinates directly from Hamiltonians $H_{1,2}$.

The crucial observation is that these variables of separation for the Neumann system coincide with the well-known variables of separation for the Chaplygin system which describes the dynamics of a rigid body moving by inertia in an infinitely extended ideal incompressible fluid \cite{ch03}. In the Chaplygin case the second integral of motion is a  polynomial in the momenta  of degree four  in contrast with the Neumann system where both integrals of motion  $H_{1,2}$ are the second order polynomials in momenta.

Then we identify  canonical transformation of the elliptic coordinates to the  Chaplygin variables with the special B\"{a}cklund transformation, which can be easily described using $2\times 2$ Lax representation for the Neumann system. It allows us to  obtain similar B\"{a}cklund transformations for other curvilinear coordinate systems and to prove that known variables of separation for the system with quartic potential \cite{rom95}, for the H\'{e}non-Heiles system \cite{rav93} and for the Kowalevski top \cite{ts13a,ts10k} are the standard curvilinear coordinates (elliptic, parabolic) after the B\"{a}cklund transformations.

This paper is organized as follows. In Section 2 we will calculate two different families of variables of separation for the Neumann system in the framework of the bi-Hamiltonian geometry. Section 3 contains the main results on simultaneous separation for the Neumann and Chaplygin systems, whereas in Section 4  other curvilinear coordinates and their B\"{a}cklund transformations are considered.

\subsection{Separation of variables in the Hamilton-Jacobi equation}
Let us consider an integrable Hamiltonian system on the phase space $M$ with coordinates $x=(x_1,\ldots,, x_n)$ and $p_x=(p_{x_1}, \ldots  p_{x_n} )$. It consists of the necessary number of  in\-de\-pen\-dent functions   $H_1,\ldots H_n$  in the involution
\[\{H_i,H_j\}=0\,,\qquad i,j=1,\ldots,n\,.\]
 Canonical coordinates $q=(q_1,\ldots,, q_n)$ and $p=(p_1, \ldots  p_n)$ are  variables of separation if
the stationary Hamilton-Jacobi equations
\[H_i\left(q,\dfrac{\partial S}{\partial q}\right)=0
\]
can be collectively solved by the additive complete integral
\[ S(q_1,\ldots,q_n;\alpha_1,\ldots,\alpha_n)= \sum_{i=1}^n
S_i(q_i,\alpha_1,\ldots,\alpha_n)\,,\qquad
\det\left\|\dfrac{\partial ^2 S}{\partial q_i\partial
\alpha_j}\right\|\neq 0\,,
\]
where $\alpha_i$ are values of $H_i$ \cite{jac66}. In this case we can find solutions $q_i=q_i(t,\alpha,\beta)$ and $p_i=p_i(t,\alpha,\beta)$
 of the initial Hamilton equations of motion   using  the Jacobi equations
\[
\beta_i=-\dfrac{\partial S(q,\alpha)}{\partial \alpha_i},\qquad
p_i=\dfrac{\partial S(q,\alpha)}{\partial q_i}\,,\qquad i=1,\ldots,n.
\]
Second Jacobi equations
\[
p_i=\dfrac{\partial S}{\partial q_i}=\dfrac{\partial S_i(q_i,\alpha_1,\ldots,\alpha_n)}{\partial
q_i}\,,
\]
and their more symmetric forms
\[ \Phi_i(q_i,p_i,\,H_1,\ldots,H_n)=0\,,\qquad i=1,\ldots,n\,, \]
usually call the {separated relations} \cite{skl95}.

In \cite{jac66} Jacobi wrote: ``The main difficulty in integrating a given differential equation
lies in introducing convenient variables, which there is no rule for finding.
Therefore, we must travel the reverse path and after finding some notable substitution,
look for problems to which it can be successfully applied."\\
 Namely,    after finding notable canonical variables
 \[  q_i=q_i(x_1,\ldots,q_n,p_{x_1},\ldots,p_{x_n})\,,\qquad\mbox{and}\qquad  p_i=p_i(x_1,\ldots,q_n,p_{x_1},\ldots,p_{x_n})\]
we can  substitute these functions into the other separated relations
\[
 \tilde{\Phi}_i(q_i,p_i,\,\tilde{H}_1,\ldots,\tilde{H}_n)=0\,,\qquad i=1,\ldots,n\,,
\]
and solve these relations with respect to $\tilde{H}_i(x,p_x)$.   It is easy to see that  the Hamilton-Jacobi equations for initial Hamiltonians $H_1,\ldots,H_n$ and new Hamiltonians $\tilde{H}_1,\ldots,\tilde{H}_n$  are simultaneously separable in variables  $q$ and $p$.

For instance,  standard definition of the  elliptic coordinates on the sphere allows us to construct an infinite family of potentials $V_j(x)$, which are simultaneously separable with the Neumann potential \cite{stef85}.  More complicated example of simultaneously separable integrable systems may be found in \cite{kuz02s}.

So, the Jacobi theory consists of the direct methods of finding  variables of separation $q$ and $p$ as functions on initial physical variables and  of the inverse methods of finding separated relations, generating  Hamiltonians $\tilde{H}_i$ that have a physical meaning in physical variables $x$ and $p_x$.

\section{Direct method}
\setcounter{equation}{0}

Let us start with a unit sphere $\mathbb S^2$ embedded in the three dimensional space $\mathbb R^3$ with Cartesian coordinates $x=(x_1, x_2, x_3)$. The corresponding vector of momenta is $p=(p_1,p_2,p_3)$.

On the six-dimensional phase space $T^*\mathbb R^3$ equipped with canonical Poisson brackets
\bq\label{can-br1}
\{x_i,p_i\}=\delta_{ij}\,,
\eq
we introduce the angular momentum vector
\[J=x\times p\,,\]
and integrals of motion for the Neumann system
\ben
H_1&=&J_1^2+J_2^2+J_3^2+a_1x_1^2+a_2x_2^2+a_3x_3^2\,,\nn\\
\label{neu-int}\\
 H_2&=&a_1J_1^2+a_2J_2^2+a_3J_3^2-a_2a_3x_1^2-a_1a_3x_2^2-a_1a_2x_3^2\,.\nn
\en
In order to realize the motion of a particle on the sphere $\mathbb S^2$ we have to impose two constraints
\[
F_1=x_1^2+x_2^2+x_3^2=1\,,\qquad F_2=x_1p_1+x_2p_2+x_3p_3=0\,,
\]
which allows us to get the desired phase space $M=T^*\mathbb S^2$ equipped with the Dirac-Poisson bracket defined by
\[
 \{f, g\}_D = \{f, g\} - \dfrac{\{F_1, f\}\{F_2, g\} - \{F_2, f\}\{F_1, g\}}{\{F_1, F_2\}}.
\]
The Dirac-Poisson brackets of the coordinate functions are:
\bq\label{d-poi}
\{x_i,x_j\}_D=0\,,\qquad \{x_i,p_j\}_D=\delta_{ij}-x_ix_j\,,\qquad \{p_j,p_j\}_D=p_ix_j-x_ip_j\,.
\eq
Note that the Dirac-Poisson structure and the Hamiltonian flow $\dot{z}=\{H_1,z\}_D$ are defined not only on $T^*{\mathbb S}^2$, but on the whole $\mathbb R^6$ without $x = 0$.

In order to identify the Dirac-Poisson bracket with a canonical Poisson bracket on the cotangent bundle $T^*{\mathbb S} ^2$ we will use the standard spherical coordinate system and the conjugated momenta
\bq\label{sph-coord}
\begin{array}{lll}
x_1 =\sin\phi\sin\theta,\qquad& x_2 = \cos\phi\sin\theta,\qquad & x_3 =\cos\theta\,,\\
\\
J_1 =\dfrac{\sin\phi\cos\theta}{\sin\theta}\,p_\phi-\cos\phi\,p_\theta\,,\qquad&
J_2 =\dfrac{\cos\phi\cos\theta}{\sin\theta}\,p_\phi+\sin\phi\,p_\theta\,,
\qquad& J_3 = -p_\phi\,,
\end{array}
\eq
so that
\bq\label{can-br2}
\{\phi,\theta\}_D=\{p_\phi,p_\theta\}_D=0\,,\quad \{\phi,p_\phi\}_D=\{\theta,p_\theta\}_D=1\,.
\eq
In these coordinates integrals of motion have the following form
\ben
H_1&=&\dfrac{p_\phi^2}{\sin^2\theta}+p_\theta^2+(a_1\sin^2\phi+a_2\cos^2\phi)\sin^2\theta+a_3\cos^2\theta,\nn\\
\label{neu-int2}\\
H_2&=&\left(
\frac{a_1\sin^2\phi+a_2\cos^2\phi}{\tan^2\theta}+a_3
\right)p_\phi^2-\frac{(a_1-a_2)\sin2\phi}{\tan\theta}p_\phi p_\theta\nn\\
& +&(a_1\cos^2\phi+a_2\sin^2\phi)(p_\theta^2-a_3\sin^2\theta)-a_1a_2\cos^2\theta\,.
\nn
\en
Recall that every Poisson bracket on a manifold $M$, dim $M=k$, with coordinates $z_1,\ldots,z_k$ is defined by the Poisson bivector $P$ by the rule
\[
\{f,g\}=\sum_{ij=1}^{k} P_{ij} \dfrac{\partial f}{\partial z_i}\dfrac{\partial g}{\partial z_j}\,.
\]
In our case canonical brackets (\ref{can-br1}) and (\ref{can-br2}) are defined by the canonical Poisson bivector
\bq\label{can-p}
P=\left(
 \begin{array}{cc}
 0 & I \\
 -I & 0 \\
 \end{array}
\right)
\eq
where $I$ is a unit matrix.

\subsection{Second Poisson brackets}
Bi-Hamiltonian manifold $M$ is a smooth manifold endowed
with a pair the Poisson bivectors $P$ and $P'$
\bq\label{m-eq2}
[ P,P] =0\,,\qquad [P',P']=0\,,
\eq
which are compatible to each other
\bq\label{m-eq1}
[P,P']=0.
\eq
Here $[ .,.] $ is a Schouten bracket defined by
\[
[ A,B] _{ijk}=-\sum\limits_{m=1}^{k}\left(B_{mk}\dfrac{\partial A_{ij}}{\partial z_m}
+A_{mk} \dfrac{\partial B_{ij}}{\partial z_m}+\mathrm{cycle}(i,j,k)\right)\,.
\]
Integrable systems on $M$ are bi-integrable if the corresponding integrals of motion $H_1,\ldots,H_n$ are in the bi-involution
\bq\label{bi-inv}
\{H_i,H_j\}=\{H_i,H_j\}'=0\,,\qquad i,j=1,\ldots,n,
\eq
with respect to the pair of compatible Poisson brackets $\{.,.\}$ and $\{.,.\}'$ defined by $P$ and $P'$.

In order to find the second bivector $P'$ on $T^*\mathbb S^2$ we can take integrals of motion $H_{1,2}$ (\ref{neu-int2})
and view (\ref{m-eq1}) and (\ref{bi-inv}) as equations to determine $P'$. Obviously enough,
the conditions on $P'$ coming from (\ref{m-eq2},\ref{m-eq1}) and (\ref{bi-inv}), in their full generality, are too difficult to be solved.

 A couple of Ans\"{a}tze which will enable us to solve them for the Neumann case are the following. Recall that for the natural Hamiltonian system
\[X=PdH\,,\qquad H=T+V=\sum g_{ij}(q)p_ip_j+V(q_1,\ldots,q_n)
\]
on the cotangent bundle $M=T^*Q$ with coordinates $z=(q_1,\ldots,q_n,p_1,\ldots,p_n)$ we can search for the Lie derivative
\[P'=\mathcal L_Y P\]
of the canonical Poisson bivector (\ref{can-p})
\[
 P=\sum_{i=1}^n \dfrac{\partial}{\partial q_i}\wedge \dfrac{\partial}{\partial p_i}\,.
 \]
 which satisfies (\ref{m-eq1}) for any $Y$.  Here $\mathcal L_Y$ is a Lie derivative of $P$ along  vector field $Y$ defined by
\[
\Bigl(\mathcal L_{Y}P\Bigr)_{ij}=\sum\limits_{k=1}^{2n}\left(Y_k\dfrac{\partial P_{ij}}{\partial z_k}
-P_{kj}\dfrac{\partial Y_i}{\partial z_k} -P_{ik}\dfrac{\partial Y_j}{\partial z_k}\right)
\]
The first Ans\"{a}tze for the vector field $Y$ is associated with the Turiel deformation of the canonical Poisson bivector on cotangent bundle \cite{tur92}. In this case entries of $Y$ are some linear functions in momenta
\bq\label{1-anz}
Y_{i}=\sum_{j=1}^n L_{ij}(q)p_j\,,\qquad i=1,\ldots,2n.
\eq
The second Ans\"{a}tze is a natural generalisation of the previous one
\bq\label{2-anz}
Y_i=\sum_{jk=1}^n \Pi_{ijk}(q)p_jp_k+\sum_{j=1}^n \Lambda_{ij}(q)p_j\,,\qquad i=1,\ldots,2n.
\eq
Some examples of the corresponding Poisson bivectors and bi-integrable systems on $T^*\mathbb S^2$ are discussed in \cite{ts11,ts11s}.

 Below we use spherical coordinates $q=(\phi,\theta)$ and the corresponding momenta $p=(p_\phi,p_\theta)$ for the Neumann system. In these variables the first solution of (\ref{m-eq2}) and (\ref{bi-inv}) looks like
\bq\label{neu-sol1}
Y^{(1)}=\left(
 \begin{array}{c} 0\\0\\
 (a_1 \cos^2\phi+a_2\sin^2\phi)\,p_\phi -\frac{b\sin 2\phi\,{\sin2\theta}}2\,p_\theta\\
 \\
 -\frac{b\sin 2\phi\,\cos\theta}{\sin\theta}\,p_\phi + \Bigl(a_3\sin^2\theta+(a_1\sin^2\phi+a_2\cos^2\phi)\cos^2\theta\Bigr)\,p_\theta
 \end{array}
 \right)\,,
\eq
where $b=(a_2-a_1)/2$. The corresponding Poisson bivector is discussed in \cite{ped04}.

The second solution has the following form
\bq\label{neu-sol2}
Y^{(2)}=\left(
 \begin{array}{c}
 Y_T \\
 Y_V \\
 \end{array}
 \right)
\,,\eq
where
\[ Y_T=
\left(
 \begin{array}{c}
 \dfrac{2p_\phi p_\theta}{\tan2\theta} \\ \\
 \cot\theta p_\phi^2-\tan\theta p_\theta^2 \\
 \end{array}
 \right)
\qquad
Y_V= -b\left(
 \begin{array}{c}
 \cos2\phi p_\phi+\sin2\phi\tan\theta p_\theta \\ \\
 \sin2\phi\cot\theta p_\phi -\cos2\phi p_\theta\\
 \end{array}
 \right)\,.
\]
This second trivial deformation $P'_{2}=\mathcal L_{Y^{(2)}}\,P$ of the canonical bivector $P$ on $T^*\mathbb S^2$ was obtained in \cite{ts10a} as a solution of the same equations (\ref{m-eq2}) and (\ref{bi-inv}) for the Chaplygin system.

\subsection{Variables of separation}

In the previous section we found two solutions $P'_{1,2}=\mathcal L_{Y_{1,2}}\,P$ of equations (\ref{m-eq1}) and (\ref{bi-inv}). According to \cite{ped04,mag97} the desired coordinates of separation are eigenvalues $\lambda_{i}$ of the recursion operator
\[
N=P'P^{-1}.
\]
For the Neumann system we have two distinct eigenvalues for both solutions of (\ref{m-eq1}) and (\ref{bi-inv}). We begin this section by stating, for the sake of completeness, the well-known definitions of the elliptic coordinates on $T^*\mathbb S^2$.

For the first vector field $Y^{(1)}$ (\ref{neu-sol1}) the coordinates of separation satisfy the equation
\bq\label{q12-neu}
e(\lambda)\equiv\dfrac{x_1^2}{\lambda-a_1}+\dfrac{x_2^2}{\lambda-a_2}+\dfrac{x_3^2}{\lambda-a_3}=0\,,
\qquad \mbox{with}\qquad x_1^2+x_2^2+x_3^2=1\,,
\eq
whereas their conjugate momenta $p_{u_{1,2}}$ are given by  values of the function
\bq\label{p12-neu}
h(\lambda)=\dfrac{1}{2}\left(\dfrac{p_1 x_1}{\lambda-a_1}+\dfrac{p_2x_2}{\lambda-a_2}+\dfrac{p_3x_3}{\lambda-a_3}\right)\,,\qquad
\eq
for $\lambda=u_{1,2}$.  Inverse transformation reads as
\bq\label{ell-coord}
\begin{array}{l}
x_i=\sqrt{\dfrac{(u_1-a_i)(u_2-a_i)}{(a_j-a_i)(a_k-a_i)}},\qquad i\neq j\neq k\,,\qquad p=J\times x\\
\\
J_i=\dfrac{2\varepsilon_{ijk}x_jx_k(a_j-a_k)}{u_1-u_2}\Bigl((a_i-u_1)p_{u_1}-(a_i-u_2)p_{u_2}\Bigr)\,,
\end{array}
\eq
where $\varepsilon_{ijk}$ is a completely antisymmetric tensor.

These coordinates were introduced by Neumann \cite{neu59} and generalized by Moser \cite{mos80-2} for arbitrary $n$. Like the elliptic coordinates in $\mathbb R^2$, the
elliptic coordinates on $\mathbb S^2$ are also orthogonal and only locally defined. They
take values in the intervals
\[a_1 < u_1 < a_2 < u_2 <a_3\]
Substituting these well-known defining relations for $x_i$ and $J_i$ (\ref{ell-coord}) in $H_{1,2}$ one gets separated relations
\bq\label{nsep-rel}
\Phi(\lambda,\mu)= 4(\lambda-a_1)(\lambda-a_2)(\lambda-a_3)\mu^2+\lambda^2+\lambda(H_1-a_1-a_2-a_3)-H_2=0\,,
\eq
for $\lambda=u_{1,2}$ and $\mu=p_{u_{1,2}}$.

For the second vector field $Y^{(2)}$ (\ref{neu-sol2}) the eigenvalues $s_{1,2}$ of recursion operator $N_2$ satisfy the equation
\bq\label{q12-ch}
\hat{f}(\lambda)\equiv \dfrac{1}{b^2-\lambda^2}\left(\lambda^2-\left(\dfrac{J_1^2+J_2^2}{x_3^2}\right)\lambda
+\dfrac{b(J_1^2-J_2^2)}{x_3^2}-b^2\right)=0
\eq
and their conjugate momenta $p_{s_{1,2}}$ are given by  values of the function
\bq\label{p12-ch}
 \hat{h}(\lambda)=\dfrac{ (x_2J_1-x_1J_2)\lambda
 -(x_2J_1+x_1J_2){b} }{2x_3(b^2-\lambda^2)}
\eq
for $\lambda=s_{1,2}$. Coordinates $s_{1,2}$ take values in the intervals
\[ s_1<b<s_2\,,\qquad b=\dfrac{a_2-a_1}{2}\,,\]
and the inverse transformation is
\ben
J_3&=&-\dfrac{x_1J_1+x_2J_2}{x_3}\,,\qquad x_3=\sqrt{1-x_1^2-x_2^2\,}\,,\nn\\
\nn\\
J_1&=&x_3\sqrt{\dfrac{(b+s_1)(s_2+b)}{2b}\,}\,,\qquad
J_2=x_3\sqrt{\dfrac{(b-s_1)(s_2-b)}{2b}\,}\,,\nn\\
\nn\\
x_1&=&\sqrt{\dfrac{2(s_2-b)(b-s_1)}{b(s_2-s_1)^2}}\bigl(p_{s_1}(s_1+b)-p_{s_2}(s_2+b)\bigr)\,,\nn\\
\nn\\
x_2&=&\sqrt{\dfrac{2(s_2+b)(b+s_1)}{b(s_2-s_1)^2}}\bigl(p_{s_1}(s_1-b)-p_{s_2}(s_2-b)\bigr)\,.\nn
\en
 Substituting variables $x_i$ and $J_i$ in the Neumann integrals of motion $H_{1,2}$ one gets the same separated equations
$\Phi(\lambda,\mu)=0$ (\ref{nsep-rel}) at
\[\lambda=-s_{1,2}+\dfrac{a_1+a_2}{2}\,,\qquad\mbox{and}\qquad \mu=p_{s_{1,2}}.\]
In \cite{ch03} Chaplygin showed that the Hamilton-Jacobi equation of the system defined by the following integrals of motion \bq\label{ham-ch}
\tilde{H}_1=J_1^2+J_2^2+2J_3^2-2b(x_1^2-x_2^2)\,,\qquad \tilde{H}_2=\left(J_1^2-J_2^2-2bx_3^2\right)^2+4J_1^2J_2^2
\eq
is additively separable at the same  coordinates
\[s_{1,2}=\dfrac{J_1^2+J_2^2\pm\sqrt{\tilde{H}_2}}{2x_3^2}\,.\]
It allows us to say that coordinates $s_{1,2}$  are  separation coordinates  for the Neumann and Chaplygin systems simultaneously.

\section{Inverse method}
\setcounter{equation}{0}

Before applying the notable substitution from the previous section to other problems, let us make some comments on their relation with the well-known elliptic coordinates. Of course, on the cotangent bundle $T^*\mathbb S^2$ there is a canonical transformation relating elliptic coordinates $u_{1,2}$ and their momenta $p_{v_{1,2}}$ with the Chaplygin coordinates $s_{1,2}$ and momenta $p_{s_{1,2}}$.  This transformation can be defined explicitly, for instance
\[
\begin{array}{lcl}
u_1+u_2&=&\frac{4(a_2-a_3-b)\bigl( (b^2-s_1^2)p_{s_1}^2- (b^2-s_2^2)p_{s_2}^2\bigr)}{s_1-s_2}+\frac{8(b^2-s_1^2)(b^2-s_2^2)p_{s_1}p_{s_2}}{(s_1-s_2)^2} \\
 \\
&-&\frac{4(b^2-s_1s_2)\bigl((b^2-s_1^2)p_{s_1}^2+(b^2-s_2^2)p_{s_2}^2 \bigr)}{(s_1-s_2)^2}+2(a_2-b)\,, \\
 \\
u_1u_2&=&\frac{4\bigl(a_2^2-a_2a_3-2a_2b+a_3b\bigr)\bigl( (b^2-s_1^2)p_{s_1}^2- (b^2-s_2^2)p_{s_2}^2\bigr)}{s_1-s_2}+\frac{8a_3(b^2-s_1^2)(b^2-s_2^2)p_{s_1}p_{s_2}}{(s_1-s_2)^2} \\
 \\
&-&\frac{4a_3(b^2-s_1s_2)\bigl((b^2-s_1^2)p_{s_1}^2+(b^2-s_2^2)p_{s_2}^2 \bigr)}{(s_1-s_2)^2}+2a_2(a_2-b)\,.
\end{array}
\]
In fact this  transformation is the special B\"{a}cklund  transformation of the elliptic coordinates for the Neumann system.

\subsection{ B\"{a}cklund  transformations}
The B\"{a}cklund transformations  arose as certain transformations between surfaces in 1875.  In the modern  theory of integrable PDEs, they are usually interpreted as:
\begin{itemize}
  \item relations between solutions of the same PDE (auto-BT)
  \item relations between solutions of  different PDEs (hetero-BT)
  \item differential-difference equations, which commutativity leads to the systems of difference equations.
\end{itemize}
All the necessary references may be found in the recent books \cite{rod02,bob09,li14}.

For finite dimensional integrable systems B\"{a}cklund  transformations  are {canonical transformations} of coordinates
\bq\label{bt-coord}
(u,p_u)\to (v,p_v)\,,\qquad \{u_i,p_{u_j}\}=\{v_i,p_{v_j}\}=\delta_{ij}\,,
\eq
which preserve  form of the Hamiltonians $H_1,\ldots,H_n$ \cite{stef82}.

  If $u$-variables and $v$-variables (\ref{bt-coord}) are variables of separation, these  B\"{a}cklund  transformations preserve the form of the separated relations
\[\Phi(u_i,p_{u_i},\,H_1,\ldots,H_n)=0 \to \Phi(v_i,p_{v_i},\,H_1,\ldots,H_n)=0\,,\]
i.e. they change coordinates on an algebraic invariant manifold without changing the manifold itself \cite{kuz02}. In this case B\"{a}cklund  transformations are  relations between common additive solutions
\[ S(q_1,\ldots,q_n;\alpha_1,\ldots,\alpha_n)= \sum_{i=1}^n
S_i(q_i,\alpha_1,\ldots,\alpha_n)\,,\qquad q=u,v\,\]
of a couple of the same PDEs
 \bq
 \label{Eq-HJt} H_i\left(q,\dfrac{\partial S}{\partial q}\right)=\alpha_i\,,\qquad q=u,v\,.
\eq
Thus, in this case we have finite dimensional analog of the auto B\"{a}cklund  transformations.

Second standard point of view suggests to consider variables $(v,p_v)$ as the
variables $(u,p_u)$, but computed at the next time step, so B\"{a}cklund  transformations (\ref{bt-coord})
 are  discretisations of the continuous model, see  \cite{bps13,kuz02,zul13} and the references within.

 We want to discuss the third standard  point of view and to consider  hetero  B\"{a}cklund  transformations, which are
  relations between additive solutions of two different systems of PDEs
   \bq
 \label{Eq-HJt-2} H_i\left(u,\dfrac{\partial S}{\partial u}\right)=0\,,\qquad\mbox{and}\qquad \tilde{H}_i\left(v,\dfrac{\partial \tilde{S}}{\partial v}\right)=0\,.
\eq
 Namely, substituting variables $ (v,p_v)$ into the separated relations
\[\tilde{\Phi}(v_i,p_{v_i},\,\tilde{H}_1,\ldots,\tilde{H}_n)=0  \]
we can get  an integrable system with Hamiltonians $\tilde{H}_k$, which do not separable in the initial coordinates $ (u, p_u) $þ As usual,  hetero  B\"{a}cklund  transformations are relations between equations rather then relations between their solutions

\subsection{Neumann and Chaplygin systems}
Let us take $2\times 2$ Lax matrix for the Neumann system
\bq\label{lax-1}
L(\lambda)
=
\left(
 \begin{array}{cc}\displaystyle
 \frac{1}{2} \sum_{k=1}^{n}\frac{x_kp_k}{\lambda-a_k} &\displaystyle \sum_{k=1}^{n}\frac{x_k^2}{\lambda-a_k} \\ \\
 \displaystyle - \frac{1}{4} \left(1+\sum_{i=1}^{n}\frac{p_k^2}{\lambda-a_k}\right) &\displaystyle - \frac{1}{2} \sum_{k=1}^{n}\frac{x_kp_k}{\lambda-a_k}\\
 \end{array}
 \right)\,,
\eq
for instance see \cite{harn88}. A separation of variables is recovered in this formalism noticing that zeroes $u_i$, $i=1,\ldots,n-1$ of the matrix entry
\[
e(\lambda)=\sum_{k=1}^{n}\frac{x_k^2}{\lambda-a_k}=\dfrac{\prod_{i=1}^{n-1} (\lambda-u_i)}{\prod_{k=1}^n (\lambda-a_k)}
\]
define the elliptic (spheroconical) coordinates, and their conjugate momenta $p_{u_i}$ are given by  the values of the rational function $L_{11}(\lambda)$ for $\lambda=u_i$.

\begin{rem}
According to \cite{ts04} function $e(\lambda)$ completely defines this Lax matrix
\bq\label{lax-e}
L(\lambda)=\left(
 \begin{array}{cc}
 \dfrac{ \{H_0,e(\lambda)\}_D}{2} & e(\lambda) \\ \\
 -\dfrac{1}{4}-\dfrac{\{H_0,\{H_0,e(\lambda)\}_D\}_D}{2}+eH_0 & -\dfrac{ \{H_0,e(\lambda)\}_D}{2} \\
 \end{array}
 \right)\,,
\eq
where $\{.,.\}_D$ is a canonical Poisson bracket on $T^*\mathbb S^ {n-1} $ (\ref{d-poi})
\[\{u_i,p_{u_k}\}_D=\delta_{ik}\,,\qquad \{u_i,u_k\}_D= \{p_{u_i},p_{u_k}\}_D=0\]
 and
\[
{H}_0=\sum_{i=1}^{n-1} \left. \mbox{Res}\right|_{\lambda=u_i}e^{-1}(\lambda)\,p^2_{u_i}\,.
\]
\end{rem}

We are now ready to introduce a second family of the variables of separation.
\begin{prop}
Let us make the similarity transformation
\bq\label{sim-tr}
\hat{L}=VLV^{-1}
\eq
 with matrix
 \[ V=\left(
 \begin{array}{cc}
 {L}_{12} & 0 \\
 4\bigl({L}_{11}-\hat{L}_{11}(\lambda)\bigr )& 4{L}_{12}\\
 \end{array}
 \right)\,,\]
where
\[
\qquad
\hat{L}_{11}(\lambda)=\dfrac{a_n-\lambda}{2}\,\bigl\{\hat{H}_0,e(\lambda)\bigr\}_D\qquad \mbox{and}\qquad
\hat{H}_0=\sum_{i=1}^{n-1} \dfrac{\left. \mbox{Res}\right|_{\lambda=u_i}e^{-1}(\lambda)}{a_n-u_i}\,p^2_{u_i}\,.
\]
The Lax matrix $\hat{L}(\lambda)$ has two off-diagonal elements with $(n-1)$ zeroes
 \bq\label{uv-coo}
\hat{L}_{12}(\lambda)=\dfrac{\prod_{i=1}^{n-1} (\lambda-u_i)}{4\prod_{k=1}^n (\lambda-a_k)}\quad\mbox{and}\quad
\hat{L}_{21}(\lambda)=-\dfrac{\prod_{i=1}^{n-1} (\lambda-v_i)}{\prod_{k=1}^{n-1} (\lambda-a_k)}\,,
\eq
which are separation coordinates for the Neumann system.
\end{prop}

It is easy to see that coordinates $u_i,\,i=1,\ldots,n-1$ are standard elliptic (spheroconical) coordinates on $\mathbb S^{n-1}$, whereas coordinates $v_i,\,i=1,\ldots,n-1$ are some new variables of separation for the Neumann system.
Indeed, the corresponding momenta $p_{u_i}$ and $p_{v_i}$ are the eigenvalues of the Lax matrix $\hat{L}$ for $\lambda=u_i$ and $\lambda=v_i$ and, therefore, each pair of canonical variables $u_i,\,p_{u_i}$ and $v_i,\,p_{v_i}$ satisfy the common separated equation
 \bq\label{ell-c}
 C:\qquad \mu^2+\mbox{det}{L}(\lambda)=0\,, \qquad \lambda=u_i,\,\mu=p_{u_i}\quad \mbox{or}\quad \lambda=v_i,\,\mu=p_{v_i}\,, \eq
which defines the genus $(n-1)$ hyperelliptic curve $C$.

At $n=3$ this standard Lax matrix (\ref{sim-tr}) looks like
\[
\hat{L}=\left(
 \begin{array}{cc}
 \dfrac{(p_1x_3-p_3x_1)x_1}{2x_3(\lambda-a_1)}+\dfrac{(p_2x_3-p_3x_2)x_2}{2x_3(\lambda-a_2)} & \displaystyle \dfrac{1}{4}\left(\sum_{i=1}^{n}\frac{x_i^2}{\lambda-a_i} \right)\\ \\
 -1-\dfrac{(p_1x_3-p_3x_1)^2}{x_3^2(\lambda-a_1)} -\dfrac{(p_2x_3-p_3x_2)^2}{x_3^2(\lambda-a_2)} & -\dfrac{(p_1x_3-p_3x_1)x_1}{2x_3(\lambda-a_1)}+\dfrac{(p_2x_3-p_3x_2)x_2}{2x_3(\lambda-a_2)} \\
 \end{array}
 \right)\,.
\]
The crucial observation is that $v$-coordinates (\ref{uv-coo}) coincide with the Chaplygin variables of separation $s_{1,2}$ up to the shift
\bq\label{v-sph}
v_i=-s_i+\dfrac{a_1+a_2}{2}\,.
\eq
Namely, substituting $v$-variables in a couple of the equations
\bq\label{ch-c}
8(\lambda-a_1)(\lambda-a_2)\mu^2+2\lambda+\tilde{H}_1\pm\sqrt{\tilde{H}_2}-a_1-a_2=0
\eq
and solving the resulting system with respect to $\tilde{H}_{1,2}$, we obtain Chaplygin's integrals of motion (\ref{ham-ch}). The conjugated momenta $p_{s_i}$ from the Chaplygin paper \cite{ch03} coincides with the momenta $p_{v_i}$ up to canonical transformation
\[p_{s_i}=p_{v_i}+f_i(v_i)\,,\]
which changes the form of the separated relations (\ref{ch-c}).

\begin{prop}
The Hamilton-Jacobi equations for the Neumann system with Hamiltonian
\[H_1=J_1^2+J_2^2+J_3^2+a_1x_1^2+a_2x_2^2+a_3x_3^2\,,\]
and for the Chaplygin system with Hamiltonian
\[
\tilde{H}_1=J_1^2+J_2^2+2J_3^2-2b(x_1^2-x_2^2)\,,\quad b=\dfrac{a_2-a_1}{2}
\]
are simultaneously separable in $v$-variables, which can be obtained from the standard elliptic coordinated on the sphere using  B\"{a}cklund transformation.
\end{prop}
At $n>3$ we can get $v$-variables from the elliptic variables as well, but we  do not know suitable separated relations, generating  physically interesting systems on $T^*\mathbb S^{n-1}$.

\begin{rem}
Using elliptic coordinates, we can get separable potentials $V (x) $ related to each other by the recurrence relation \cite{stef85}.  It would be interesting to obtain similar relations for $v$-variables. For instance, let us consider an integrable system on the sphere with quartic potential separable in elliptic coordinates associated with the Lax matrix
\[
\tilde{L}(\lambda)= L(\lambda)+a\left(
 \begin{array}{cc}
 0 & 0 \\ \\
 \lambda-\sum_{k=1}^n a_kx_k^2 &0 \\
 \end{array}
 \right)\,,\qquad a\in \mathbb R\,.
\]
Here $L(\lambda)$ is given by (\ref{lax-1}) and $a$ is an arbitrary number.  In order to get the necessary number of zeroes in both off-diagonal elements of the Lax matrix we can make the similarity transformation (\ref{sim-tr}) with
\[
\hat{L}_{11}=\dfrac{a_n-\lambda}{2}\,\bigl\{\hat{H}_0,e\bigr\}_D+\sqrt{a}\,(\lambda-a_n)\,e(\lambda)\,.
\]
At $n=3$ the corresponding spectral curve
\[
C:\qquad \mu^2(\lambda-a_1)(\lambda-a_2)(\lambda-a_3)-a\lambda^3-\left(\dfrac{1}{4}+a(a_1+a_2+a_3)\right)\lambda^2-H_1\lambda+H_2=0
\]
is a genus-2 hyperelliptic curve at any value of $a$. At $a=0$ we have the separated relations (\ref{ch-c}), which yield integrals of motion for the Chaplygin system. At $a\neq0$ we do not know their suitable generalizations, which allow us to get generalization of the Chaplygin system.

According to \cite{ts04sl} at $n=3$ this integrable system with quartic potential on the sphere is equivalent to the Steklov-Lyapunov integrable case of  rigid body motion. In this case $u$-variables are the so-called K\"{o}tter variables, while the second $v$-variables were found in \cite{ts12} using the direct bi-Hamiltonian method discussed in the previous Section.
\end{rem}

 \subsection{The r-matrix structures}
 Roughly speaking, after the similarity transformation (\ref{sim-tr}) zeroes of the two off-diagonal matrix elements
\[\hat{L}_{12}(\lambda=u_i)=0\qquad \mbox{and}\qquad\hat{L}_{21}(\lambda=v_i)=0
\]
form two families of separation variables with different properties. Below we want to discuss other differences between Lax matrices $L(\lambda)$ (\ref{lax-1}) and $\hat{L}(\lambda)$ (\ref{sim-tr}).

Let us rewrite the Poisson brackets of the Lax matrix entries in a specific algebraic form involving  classical $r$-matrix depending on parameters $\lambda$ and $\mu$
\bq\label{rp}
\{L_1(\lambda),\,L_2(\mu)\}=[r_{12}(\lambda,\mu),L_1(\lambda)]-[r_{21}(\lambda,\mu),L_2(\mu)] \,.
\eq
Here we use the familiar notation for the tensor product of $L$ and unit matrix $I$
\[L_{1}(\lambda)=L(\lambda)\otimes \mathrm I\,,\qquad L_2(\mu)=\mathrm I\otimes L(\mu),\qquad r_{21}(\lambda,\mu)=\Pi r_{12}(\mu,\lambda) \Pi\,,\]
and $\Pi$ is the permutation operator: ${\Pi}x\otimes y =y\otimes x$,\ $\forall x,y$ \cite{skl95}.

Canonical Poisson brackets (\ref{can-br1}) on $T^* \mathbb R^3$ between the entries of the first Lax matrix $L(\lambda)$ (\ref{lax-1}) for the Neumann system have the form (\ref{rp}) at
\bq\label{r-1}
r_{12}(\lambda,\mu)=\dfrac{1}{\lambda-\mu}
\left(\begin{array}{cccc}
 1 & 0 & 0 & 0 \\
 0 & 0 & 1 & 0 \\
 0 & 1 & 0 & 0 \\
 0 & 0 & 0 & 1
\end{array}\right)\,.
\eq
The Dirac reduction procedure shifts this $r$-matrix by dynamical term
\bq\label{r-2}
r^D_{12}(\lambda,\mu)=\dfrac{1}{\lambda-\mu}
\left(\begin{array}{cccc}
 1 & 0 & 0 & 0 \\
 0 & 0 & 1 & 0 \\
 0 & 1 & 0 & 0 \\
 0 & 0 & 0 & 1
\end{array}\right)
+\left(
 \begin{array}{cccc}
 0 & 0 &0 & 0 \\
 0 & 0 &L_{12}(\mu) & 0 \\
 0 & 0 & 0 & 0 \\
 -\frac{1}{4}-L_{21}(\mu) & 0 & 0 & 0
 \end{array}
 \right)\,,
\eq
which depends on dynamical variables via entries of the Lax matrix (\ref{lax-1}).

Canonical Poisson brackets (\ref{can-br1}) on $T^* \mathbb R^3$ between the entries of the second Lax matrix $\hat{L}(\lambda)$ (\ref{sim-tr}) have the form (\ref{rp}) at
\bq\label{rh-1}
\hat{r}_{12}(\lambda,\mu)=\dfrac{1}{\lambda-\mu}
\left(\begin{array}{cccc}
 1 & 0 & 0 & 0 \\
 0 & 0 & \frac{\lambda-a_n}{\mu-a_n} & 0 \\
 0 & 1 & 0 & 0 \\
 0 & 0 & 0 & 1
\end{array}\right)\,.
\eq
In contrast with the previous case the Dirac modification of the Poisson structure does not change this $r$-matrix
\bq
\hat{r}^D_{12}(\lambda,\mu)=\hat{r}_{12}(\lambda,\mu)\,.
\eq
It is the main visible difference between $L(\lambda)$ (\ref{lax-1}) and $\hat{L}(\lambda)$ (\ref{sim-tr}). We have to underline that the classical $r$-matrix associated with the second Lax matrix is a non-dynamical matrix on $T^*\mathbb S^2$. Hence, the new separation stands a good chance to be quantized.

The second Poisson brackets, associated with  vector fields $Y^ {(1)} $ (\ref{neu-sol1}) and $Y^ {(2)} $ (\ref{neu-sol2}), also have the linear $r$-matrix form (\ref{rp}) and the corresponding classical $r$-matrices read as
\bq\label{r-3}
\hat{r}^{(1)}_{12}(\lambda,\mu)=\dfrac{1}{\lambda-\mu}
\left(\begin{array}{cccc}
 \mu & 0 & 0 & 0 \\
 0 & 0 & \mu\,\frac{\lambda-a_n}{\mu-a_n} & 0 \\
 0 & \lambda & 0 & 0 \\
 0 & 0 & 0 & \mu
\end{array}\right)
+4(a_n-\lambda)\left(
 \begin{array}{cccc}
 0 & 0 &0 & 0 \\
 0 & 0 & \hat{L}_{12}(\mu) & 0 \\
 \frac{ \rho_1}{2(a_n-\lambda)} & 0 & 0 & 0 \\ \\
 1-\hat{ L}_{21}(\mu) & 0 & 0 & 0
 \end{array}
 \right)
\eq
and
\bq\label{r-4}
\hat{r}^{(2)}_{12}(\lambda,\mu)=\dfrac{1}{\lambda-\mu}
\left(\begin{array}{cccc}
 \mu & 0 & 0 & 0 \\
 0 & 0 & \lambda\,\frac{\lambda-a_n}{\mu-a_n} & 0 \\
 0 & \mu & 0 & 0 \\
 0 & 0 & 0 & \mu
\end{array}\right)
+\left(
 \begin{array}{cccc}
 0 & 0 &\frac{\rho_2}{a_n-\mu} & \frac{1}{4(a_n-\mu)}-\hat{L}_{12}(\mu) \\
 0 & 0 & 0& 0 \\
 0 & \hat{L}_{21}(\mu) & 0 & 0 \\
 0 & 0 & 0 & 0
 \end{array}
 \right)\,.
\eq
Functions $\rho_{1,2}$ are defined by entries of $\hat{L}$:
\[
\rho_1=\left[\dfrac{\hat{L}_{11}(\lambda)}{\hat{L}_{12}(\lambda)}\right]\qquad\mbox{and}\qquad \rho_2=\left[\dfrac{(a_n-\lambda)\hat{L}_{11}(\lambda)}{\hat{L}_{21}(\lambda)}\right]\,,
\]
where$ [X /Y ]$ is a quotient of the polynomials $X$ and $Y$ in variable $\lambda$ over a field.

\section{Other pairs of simultaneously separable systems}
\setcounter{equation}{0}

In previous Section we identify Chaplygin's variables with the standard elliptic coordinates after the B\"{a}cklund transformation.   Recall, that using  elliptic coordinates on the two-dimensional sphere we can construct a family of integrable systems with integrals of motion, which are the second order polynomials in momenta  \cite{stef85}. The obtained B\"{a}cklund transformation allows us to extend this family and to consider integrable system with quadratic Hamilton function and second integral of motion, which  is a  polynomial in the momenta  of degree four. In order to generalize this single exercise to an empirical rule, we  present a sufficient set of similar facts for other curvilinear coordinate systems in $T^*\mathbb S^ {2} $ and $T^*\mathbb R^2$.

\subsection{Neumann-Rosochatius system }
It was noticed by Rosochatius\cite{ros77} that the potential given by the sum of  the inverses of the squares of the Cartesian coordinates can be added to the quadratic Neumann potential without losing its separability property. The system so obtained is customarily called the Neumann-Rosochatius system.

According to  \cite{harn88}, the $2\times 2$ Lax matrix for the Neumann-Rosochatius system is
\bq\label{lax-nr}
{L}(\lambda)=\left(
 \begin{array}{cc}\displaystyle\frac{1}{2} \sum_{k=1}^{n}\frac{x_kp_k}{\lambda-a_k}+
 \dfrac{\mathrm i}{2}\sum_{k=1}^{n} \dfrac{b_k}{\lambda-a_k} & \displaystyle \sum_{k=1}^{n}\frac{x_k^2}{\lambda-a_k} \\ \\
 \displaystyle
 \displaystyle - \frac{1}{4} \left(1+\sum_{i=1}^{n}\frac{p_k^2+b_k^2/x_k^{2}}{\lambda-a_k}\right)
 &\displaystyle -\frac{1}{2} \sum_{k=1}^{n}\frac{x_kp_k}{\lambda-a_k}+ \dfrac{\mathrm i}{2}\sum_{k=1}^{n} \dfrac{b_k}{\lambda-a_k} \\
 \end{array}
 \right)\,,\qquad \mathrm i=\sqrt{-1}\,.
\eq
We are looking for  matrix $\hat{L}=VLV^{-1}$ (\ref{sim-tr}), which satisfies the $r$-matrix equation (\ref{rp}) with the $r$-matrix given by (\ref{rh-1}). For instance, we can take
\[
\hat{L}_{11}(\lambda)=\dfrac{a_n-\lambda}{2}\,\bigl\{\hat{H}_0,e(\lambda)\bigr\}_D+\dfrac{\mathrm i}{2}\sum_{k=1}^{n-1} \dfrac{b_k}{\lambda-a_k} -\dfrac{\mathrm ib_n}{2x_n^2}\left(e(\lambda)-\frac{x_n^2}{\lambda-a_n}\right)\,.
\]
In this case, both off-diagonal elements of $\hat{L}$ have the necessary number of zeroes
\bq\label{v-nr}
\hat{L}_{12}(\lambda=u_i)=0\qquad \mbox{and}\qquad \hat{L}_{21}(\lambda=v_i)=0\,,\qquad i=1,\ldots,n-1,
\eq
which are the coordinates of separation for the Neumann-Rosochatius system associated with the spectral curve
\bq\label{c-nr}
C:\quad 4\prod_{k=1}^n(\lambda-a_k)\left(\mu^2-\mu\sum_{k=1}^n\dfrac{\mathrm i b_k}{\lambda-a_k}\right)+\lambda^{n-1}+\lambda^{n-2}H_1+\ldots+H_{n-1}=0\,.
\eq
As usual the corresponding momenta $p_{u_{1,2}} $ and $p_{v_{1,2}} $ are given by  the values of the function $\hat{L}_{11}(\lambda)$ for $\lambda=u_i$ and $\lambda=v_i$, respectively.

At $n=3$ zeroes of  both off-diagonal elements of $\hat{L}$ are the coordinates of separation for the Hamilton-Jacobi equation associated with the Hamilton function
\bq\label{ros-h}
{H}_1=J_1^2+J_2^2+J_3^2+a_1x_1^2+a_2x_2^2+a_3x_3^2+\dfrac{b^2_1}{x_1^2}+\dfrac{b^2_2}{x_2^2}
+\dfrac{b^2_3}{x_3^2}\,.
\eq
For $u$-variables the inverse transformation is given by (\ref{ell-coord}). For $v$-variables the inverse transformation is
\[\begin{array}{lcl}
x_1^2&=&\phantom{-}\frac{4\bigl((a_2-v_1)p_{v_1}-(a_2-v_2)p_{v_2}\bigr)^2(a_1-v_1)(a_1-v_2)}{(a_1-a_2)(v_1-v_2)^2}-
\frac{4\mathrm i \bigl((a_2-v_1)p_{v_1}+(a_2-v_2)p_{v_2}\bigr) b_1}{v_1-v_2}\,,\\ \\
x_2^2&=&-\frac{4\bigl((a_1-v_1)p_{v_1}-(a_1-v_2)p_{v_2}\bigr)^2(a_1-v_1)(a_1-v_2)}{(a_1-a_2)(v_1-v_2)^2}-
\frac{4\mathrm i \bigl((a_2-v_1)p_{v_1}+(a_2-v_2)p_{v_2}\bigr) b_1}{v_1-v_2}\,,\\ \\
p_1&=&
\frac{2(a_1-v_1)(a_2-v_1)}{x_1(a_1-a_2)(v_1-v_2)}\left((a_1x_1^2-a_2x_1^2-a_1+v_1)p_{v_2}-(a_1x_1^2-a_2x_1^2-a_1+v_2)p_{v_1}\right)\\ \\
&+& \mathrm i\bigl(b_1+b_2+b_3)x_1 -\frac{\mathrm ib_1}{x_1}\,,\\ \\
p_2&=&
\frac{2(a_1-v_1)(a_2-v_1)}{x_2(a_1-a_2)(v_1-v_2)}\left((a_1x_2^2-a_2x_2^2+a_2-v_1)p_{v_2}-(a_1x_2^2-a_2x_2^2+a_2-v_2)p_{v_1}\right)\\ \\
&+& \mathrm i(b_1+b_2+b_3)x_2 -\frac{\mathrm ib_2}{x_2}\,,\\ \\
x_3&=&\sqrt{1-x_1^2-x_2^2\,}\,,\qquad p_3=-\frac{x_1p_1+x_2p_2}{x_3}\,.
\end{array}
\]
\begin{rem}
It is easy to see that we can put $b_3=0$ in the Lax matrix $\hat{L}(\lambda)$ using canonical transformation
\[
J_1\to J_1+\dfrac{\mathrm i b_3x_2}{x_3}\qquad \mbox{and}\qquad J_2\to J_2-\dfrac{\mathrm i b_3x_1}{x_3}\,,
\]
because
\[
\hat{L}_{11}=\frac{x_1\left(J_2-\mathrm i b_3x_1/x_3\right)}{2x_3(\lambda-a_1)}-\frac{x_2\left(J_1+\mathrm i b_3x_2/x_3\right)}{2x_3(\lambda-a_2)}
+\frac{\mathrm i b_1}{2(\lambda-a_1)}+\frac{\mathrm i b_2}{2(\lambda-a_2)}
\]
and
 \[
\hat{L}_{21}=-1-\frac{1}{\lambda-a_1}\left( \frac{\left(J_2-\mathrm i b_3x_1/x_3\right)^2}{x_3^2}+\frac{b_1^2}{x_1^2}\right)
 -\frac{1}{\lambda-a_2}\left(\frac{\left(J_1+\mathrm i b_3x_2/x_3\right)^2}{x_3^2} +\frac{b_2^2}{x_2^2}\right)\,.
 \]
\end{rem}

Let us put $b_3=0$ and substitute variables $\lambda=v_{1,2}$ and $\mu=p_{v_{1,2}}$ into the separated equation
\ben
\tilde{C}:\quad&&\Bigl(8(\lambda-a_1)(\lambda-a_2)\mu^2+8\mathrm i\bigl( b_1(\lambda-a_2)+b_2(\lambda-a_1)\bigr)\mu+2\lambda+\tilde{H}_1+\sqrt{\tilde{H}}_2\Bigr)\nn\\
&\times&\Bigl(8(\lambda-a_1)(\lambda-a_2)\mu^2+8\mathrm i\bigl( b_1(\lambda-a_2)+b_2(\lambda-a_1)\bigr)\mu+2\lambda+\tilde{H}_1-\sqrt{\tilde{H}}_2\Bigr)\nn\\
&-&4b_4\lambda - 16b_5(\lambda-a_1)(\lambda-a_2)\mu+8\mathrm i b_5\Bigl(b_1(\lambda-a_2)+b_2(\lambda-a_1)\Bigr)=0\label{c-ch2}
\en
defining genus three non-hyperelliptic curve $\tilde{C}$. In this case solving the resulting separated relations with respect to $\tilde{H}_1$ and $\tilde{H}_2$ we obtain the following Hamilton function
\[
\tilde{H}_1=J_1^2+J_2^2+2J_3^2-\dfrac{2b_5(J_1x_2-J_2x_1)}{x_3^3}-(a_2-a_1)(x_1^2-x_2^2)+\dfrac{b_4}{x_3^2}
+(2-x_3^2)\left(\dfrac{b_1^2}{x_1^2}+\dfrac{b_2^2}{x_2^2}\right)\,.
\]
This Hamiltonian after the canonical transformation
\[J_1\to J_1+\dfrac{b_5x_2}{x_3^3}\qquad \mbox{and}\qquad J_2\to J_2-\dfrac{b_5x_1}{x_3^3}\,,\]
has a natural form, which we describe in the  following Proposition.
\begin{prop}
The Hamilton-Jacobi equations associated with Hamilton functions
\[
{H}_1=J_1^2+J_2^2+J_3^2+a_1x_1^2+a_2x_2^2+a_3x_3^2+\dfrac{b^2_1}{x_1^2}+\dfrac{b^2_2}{x_2^2}
+\dfrac{b^2_3}{x_3^2}
\]
and
\bq\label{ham-ch2}
\tilde{H}_1=J_1^2+J_2^2+2J_3^2-(a_2-a_1)(x_1^2-x_2^2)+\dfrac{b_4}{x_3^2}
+b_5^2\left(\dfrac{1}{x_3^6}-\dfrac{1}{x_3^4}\right)+(2-x_3^2)\left(\dfrac{b_1^2}{x_1^2}+\dfrac{b_2^2}{x_2^2}\right)
\eq
are simultaneously separable in $v$-variables, which can be obtained from the standard elliptic coordinated on the sphere using  B\"{a}cklund transformation.
\end{prop}

 The separated relation (\ref{c-ch2}) was found in \cite{ts13a} using the direct bi-Hamiltonian method discussed in the previous Section. Substituting $\hat{L}(\lambda)$ into the linear $r$-matrix equation (\ref{rp}) with $r$-matrices (\ref{r-3}) or (\ref{r-4}) we can recover the corresponding Poisson bivectors $P'_1$ and $P'_2$.

\subsection{Elliptic coordinates in $\mathbb R^n$}
All orthogonal separable coordinate systems can be viewed as an orthogonal
sum of certain basic coordinate systems, which can be obtained as a proper degeneration of the elliptic coordinate system introduced by Jacobi. Recall that elliptic coordinate system $u_k$ in $\mathbb R^n$ with parameters
$a_1<a_2<\cdots< a_n$ is defined through the equation
\bq\label{ell-rn}
e(\lambda)=\prod_{k=1}^n \dfrac{\lambda-u_k}{\lambda-a_k}=1+\sum_{k=1}^n\frac{x_k^2}{\lambda-a_k}\,.
\eq
A thorough discussion of its general properties as well as its use
for separation of variables in the Hamilton-Jacobi equation can be found in \cite{jac66}.

Let us consider the following well-known Lax matrix associated with a quartic potential separable in elliptic coordinates
\bq\label{lax-ell1}
L(\lambda)=\left(
 \begin{array}{cc}
 \displaystyle \frac{1}{2}\sum_{k=1}^n\frac{x_kp_k}{\lambda-a_k} - \dfrac{\mathrm i}{2}\sum_{k=1}^{n} \dfrac{b_k}{\lambda-a_k} &\displaystyle 1+\sum_{k=1}^n\frac{x_k^2}{\lambda-a_k} \\
 \displaystyle a\lambda-a\sum_{k=1}^n x_k^2 -\frac{1}{4}\sum_{k=1}^n\frac{p_k^2+b_k^2x_k^{-2}}{\lambda-a_k} &\displaystyle -\frac{1}{2}\sum_{k=1}^n\frac{x_kp_k}{\lambda-a_k}- \dfrac{\mathrm i}{2}\sum_{k=1}^{n} \dfrac{b_k}{\lambda-a_k} \\
 \end{array}
 \right)\,,
\eq
which satisfies the linear $r$-matrix equation (\ref{rp}) with the classical $r$-matrix
\bq\label{r-ell2}
r_{12}(\lambda,\mu)=\dfrac{1}{\lambda-\mu}
\left(\begin{array}{cccc}
 1 & 0 & 0 & 0 \\
 0 & 0 & 1 & 0 \\
 0 & 1 & 0 & 0 \\
 a(\lambda-\mu) & 0 & 0 & 1
\end{array}\right)\,.
\eq
After similarity transformation $\hat{L}=V{L}V^{-1}$ (\ref{sim-tr}) with
\[
\hat{L}_{11}(\lambda)=\dfrac{a_n-\lambda}{2}\,\bigl\{\hat{H}_0,e(\lambda)\bigr\}-\dfrac{\mathrm i}{2}\sum_{k=1}^{n-1} \dfrac{b_k}{\lambda-a_k} +\dfrac{\mathrm ib_n}{2x_n^2}\left(e(\lambda)-\frac{x_n^2}{\lambda-a_n}\right)\,.
\]
one gets Lax matrix $\hat{L}$ satisfying the $r$-matrix equation (\ref{rp}) with the classical $r$-matrix (\ref{rh-1}).
Here $\{.,.\}$ means the canonical Poisson bracket on $T^*\mathbb R^n$ and
\[
\hat{H}_0=\sum_{i=1}^{n} \dfrac{\left. \mbox{Res}\right|_{\lambda=u_i}e^{-1}(\lambda)}{a_n-u_i}\,p^2_{u_i}\,.
\]
For instance, at $n=2$ we have
\[
\hat{L}_{11}(\lambda)=\dfrac{x_1(p_1x_2-p_2x_1)-(\lambda-a_1)p_2}{2x_2(\lambda-a_1)}-\dfrac{\mathrm i b_1}{2(\lambda-a_1)}
+\dfrac{\mathrm ib_2}{2x_2^2}\left(1+\dfrac{x_1^2}{\lambda-a_1}\right)\,.
\]
 In this case zeroes of  both off-diagonal elements of $\hat{L}$ are the coordinates of separation for the Hamilton-Jacobi equation associated with the following Hamilton function
\bq\label{h-ell1}
H_1=p_1^2+p_2^2+4a \Bigl((x_1^2+x_2^2)^2-a_1x_1^2-a_2x_2^2\Bigr)+\dfrac{b_1^2}{x_1^2}+\dfrac{b_2^2}{x_2^2}\,.
\eq
Both families of variables of separation $(u,p_{u})$ and $(v,p_{v})$ are associated with a common spectral curve
\bq\label{c-ell1}
C:\quad 4(\lambda-a_1)(\lambda-a_2)\left(\mu^2-\mu\sum_{k=1}^{n} \dfrac{\mathrm ib_k}{\lambda-a_k}\right)+4a\lambda^3-4a(a_1+a_2)\lambda^2+\lambda H_1-H_2=0\,.
\eq
The first coordinates are elliptic coordinates on the plane $u_{1,2}$ (\ref{ell-rn}),
whereas the second $v$-coordinates are the zeroes of the second off-diagonal element
\bq\label{v-quart}
\hat{L}_{21}=4a\lambda-4a(x_1^2+x_2^2)+\dfrac{(b_2+\mathrm i p_2x_2)^2}{x_2^4}
+\dfrac{1}{\lambda-a_1}\left(
\dfrac{\bigl(\mathrm i x_2(p_1x_2-p_2x_1)-b_2x_1\bigr)^2}{x_2^4}-\dfrac{b_1^2}{x_1^2}
\right)\,.
\eq
The corresponding momenta are values of the function
\[
\hat{L}_{11}=\dfrac{\mathrm i b_2-x_2p_2}{2x_2^2}+\dfrac{-\mathrm i(b_1x_2^2-b_2x_1^2)+x_1x_2(p_1x_2-p_2x_1)}{2x_2^2(\lambda-a_1)}
\]
for $\lambda=u_{1,2}$ and $\lambda=v_{1,2}$, respectively.

The crucial observation is that these $v$-variables coincide with the variables of separation for another integrable system with quartic potential \cite{rom95} up to canonical transformation. Namely, let us take two copies of the equation
\ben\label{c-ell2}
\tilde{C}:\qquad&&\Bigl(4(\lambda-a_1)\mu^2+4\mathrm i b_1\mu-4a\lambda^2+\tilde{H}_1+\sqrt{\tilde{H}_2}\Bigr)\times\\
&&\Bigl(4(\lambda-a_1)\mu^2+4\mathrm i b_1\mu-4a\lambda^2+\tilde{H}_1-\sqrt{\tilde{H}_2}\Bigr)+4 a b_3^2 \lambda-8 a b_4 (\lambda-a_1)\mu=0\nn
\en
defining genus three non-hyperelliptic curve $\tilde{C}$, at $\lambda=v_{1,2}$ and $\mu=p_{v_{1,2}}$ and solve the resulting system of separated relations with respect to $\tilde{H}_{1,2}$. As a result, we obtain the well-known Hamiltonian
\bq\label{h-ell2}
\tilde{H}_1=p_1^2+p_2^2+a\left(
4x_1^4+3x_1^2x_2^2+\dfrac{x_2^4}{2}-8a_1x_1^2-2a_1x_2^2\right)+\dfrac{b_1^2}{x_1^2}+\dfrac{b_3^2}{x_2^2}-\dfrac{b_4^2}{x_2^6}
\eq
after canonical transformation
\[
x_2\to x_2=\dfrac{\sqrt{2x_2^2-4a_1}}{2}\,,\quad
p_2\to p_2=\dfrac{\sqrt{2x_2-4a_1}p_2}{x_2}-\dfrac{\sqrt{2x_2^2-4a_1}(\mathrm ib_2x_2^4-b_4x_2^2+2a_1b_4)}{x_2^4(2a_1-x_2^2)}\,.
\]
A suitable separated relation (\ref{c-ell2}) was found in \cite{ts11a} together with the variables of separation and bi-Hamiltonian structure, see also \cite{tt12}.

Summing up, we prove the following
\begin{prop}
The Hamilton-Jacobi equations associated with Hamilton functions
\[
H_1=p_1^2+p_2^2+4a \Bigl((x_1^2+x_2^2)^2-a_1x_1^2-a_2x_2^2\Bigr)+\dfrac{b_1^2}{x_1^2}+\dfrac{b_2^2}{x_2^2}\,.
\]
and
\[
\tilde{H}_1=p_1^2+p_2^2+a\left(
4x_1^4+3x_1^2x_2^2+\dfrac{x_2^4}{2}-8a_1x_1^2-2a_1x_2^2\right)+\dfrac{b_1^2}{x_1^2}+\dfrac{b_3^2}{x_2^2}-\dfrac{b_4^2}{x_2^6}
\]
 are simultaneously separable in the $v$-coordinates (\ref{v-quart}) obtained from elliptic coordinates by using B\"{a}cklund transformation.
\end{prop}

\subsection{Parabolic coordinates in $\mathbb R^n$}
It is possible to degenerate the elliptic coordinate system (\ref{ell-rn}) in a proper way by
letting two or more of the parameters $a_k$ coincide. Then, the ellipsoid will become
a spheroid, or even a sphere if all of the parameters coincide.

On the other hand, if we substitute new Cartesian coordinates $x'_{i}$ defined by
\[
x_i=\dfrac{x'_i}{\sqrt{a_n}}\,,\quad i=1,\ldots,n-1,\qquad x_n=\dfrac{x'_n-a_n}{\sqrt{a_n}}
\]
in (\ref{ell-rn}), let $a_2$ tend to infinity and drop the primes, then we obtain a parabolic coordinate system defined by equation
\[
e(\lambda)=\frac{\displaystyle \prod_{k=1}^n( \lambda-u_k)}{\displaystyle \prod_{j=1}^{n-1} (\lambda-a_j)}=\lambda-2x_n-\sum_{i=1}^{n-1}\frac{x_i^2}{\lambda-a_i}\,.
\]
This transformation changes the form of the Lax matrix $L(\lambda)$ (\ref{lax-ell1}), but it preserves the classical $r$-matrix (\ref{r-ell2}). It is easy to prove that the corresponding Lax matrix $\hat{L}=V{L}V^{-1}$ (\ref{sim-tr}) satisfies the $r$-matrix equation (\ref{rp}) with the classical$r$-matrix (\ref{rh-1}) at $a_n=0$.

For instance, if $n=2$ we have the following Lax matrix
 \bq\label{par-lax1}
 L(\lambda)=\left(
 \begin{array}{cc}
 \dfrac{p_2}{2}+\dfrac{p_1x_1+\mathrm ib_1}{\lambda-a_1} &\lambda-2x_2-\dfrac{x_1^2}{\lambda-a_1} \\ \\
 a\lambda^2+2ax_2\lambda+a(x_1^2+4x_2^2)+\dfrac{p_1^2+b_1^2x_2^{-2}}{4(\lambda-a_1)} &- \dfrac{p_2}{2}-\dfrac{p_1x_1-\mathrm ib_1}{\lambda-a_1} \\
 \end{array}
 \right)\,.
 \eq
Using canonical transformations we can always put $a_1=0$ without loss of generality. In this case the equation for the spectral curve of $L(\lambda)$
\bq\label{c-par1}
C:\quad \mu^2-\dfrac{\mathrm i b_1\mu}{\lambda}-a\lambda^3-4H_1+\dfrac{H_2}{\lambda}=0
\eq
contains the Hamilton function
\bq\label{hh-1}
H_1=p_1^2+p_2^2-16ax_2(x_1^2+2x_2^2)+\dfrac{b_1^2}{x_1^2}
\eq
for the well-studied H\'{e}non-Heiles system separable in parabolic coordinates \cite{fordy91}.

After the similarity transformation $\hat{L}=V{L}V^{-1}$ (\ref{sim-tr}) with
\[
\hat{L}_{11}(\lambda)=\dfrac{p_2}{2}+\dfrac{(x_1p_1+\mathrm i b_1)(\lambda-2x_2)}{2x_1^2}
\]
one gets two families of  coordinates of separation for the H\'{e}non-Heiles system defined by the Hamilton function (\ref{hh-1}). The first variables $u_{1,2}$ are parabolic coordinates on the plane, whereas the second pair of variables $v_{1,2}$ is defined by equation
\ben\label{v-hh}
\hat{L}_{21}=4a\lambda^2+\dfrac{(8ax_1^2x_2-p_1^2)\lambda}{x_1^2}&+&
4a(x_1^2+4x_2^2)+\dfrac{2p_1(p_1x_2-p_2x_1)}{x_1^2}\\ \nn\\
&-&\dfrac{2\mathrm i(\lambda p_1-2p_1x_2+p_2x_1)b_1}{x_1^3}+\dfrac{(\lambda-2x_2)b_1^2}{x_1^4}=0\,.\nn
\en
The corresponding momenta are values of the function $\hat{L}_{11}(\lambda)$ for $\lambda=u_{1,2}$ and $\lambda=v_{1,2}$, respectively.  The  proposed B\"{a}cklund transformation  is different from the B\"{a}cklund transformation for the H\'{e}non-Heiles  discussed in \cite{kuz02}.

As above the main result is that these $v$-variables coincide with variables of separation introduced in \cite{rav93} for other H\'{e}non-Heiles up to canonical transformation.  Namely, if we substitute variables $v_{1,2}$ and $p_{v_{1,2}}$ into the following equation
\bq\label{c-par2}
\tilde{C}:\quad(4\mu^2-4a\lambda^3-\tilde{H}_1-\sqrt{\tilde{H}_2})(4\mu^2-4a\lambda^3-\tilde{H}_1+\sqrt{\tilde{H}_2})+8b_2\mu+4ab_3\lambda=0
\eq
and solve these separation relations with respect to $\tilde{H}_{1,2}$, then we obtain the well-known Hamilton function
\bq\label{hh-2}
\tilde{H}_1=p_1^2+p_2^2-2ax_2(3x_1^2+16x_2^2)-\dfrac{b_2^2}{x_1^6}+\dfrac{b_3}{x_1^2}
\eq
after canonical transformation
\[
p_1\to \sqrt{2}p_1+\dfrac{ \mathrm i\sqrt{2}(\mathrm i ab_2-b_1x_1^2)}{x_1^3}\,,\qquad x_1\to\dfrac{x_1}{\sqrt{2}}\,.
\]
The separated relation (\ref{c-par2}) defines the genus-3 non-hyperelliptic curve $\tilde{C}$ which was found in \cite{ts11a} together with the variables of separation and the bi-Hamiltonian structure.

\begin{prop}
The Hamilton-Jacobi equations associated with  two H\'{e}non-Heiles systems defined by Hamilton functions
\[
H_1=p_1^2+p_2^2-16ax_2(x_1^2+2x_2^2)+\dfrac{b_1^2}{x_1^2}
\]
and
\[
\tilde{H}_1=p_1^2+p_2^2-2ax_2(3x_1^2+16x_2^2)-\dfrac{b_2^2}{x_1^6}+\dfrac{b_3}{x_1^2}
\]
 are simultaneously separable in the $v$-coordinates  (\ref{v-hh}) obtained from parabolic coordinates by using B\"{a}cklund transformation.
\end{prop}

\subsection{Kowalevski system}
Let us  consider the non-orthogonal curvilinear coordinates $u_ {1,2} $ on $\mathbb S^2$ defined through an equation
\[
e=\dfrac{(\lambda-u_1)(\lambda-u_2)}{\lambda(\lambda^2-a^2)}=\dfrac{2\sin\theta(1+\sin\phi)}{\lambda-a}+\dfrac{2\sin\theta(1-\sin\phi)}{\lambda+b}-\dfrac{1+4\sin\theta}{\lambda}\,.
\]
Starting with this function $e(\lambda)$ we can get the $2\times 2$ Lax matrix (\ref{lax-e}), which satisfies the linear $r$-matrix equation (\ref{rp}) with the classical $r$-matrices (\ref{r-1}) or (\ref{r-2}), see \cite{ts04}.

In  similar manner we can directly construct a representation of the $r$-matrix algebra
(\ref{rp}) associated with the classical $r$-matrix (\ref{rh-1}). In our case $a_n=0$ and entries of the corresponding Lax matrix read as
\ben
\hat{L}_{11}&=& \scriptstyle \frac{a\cos\phi\,p_\phi}{\lambda^2-a^2}+\frac{\tan\theta(\lambda+a\sin\phi)p_\theta}{\lambda^2-a^2}-\frac{\mathrm i \sqrt{b_1-b_2}}{2(\lambda-a)}-\frac{\mathrm i\sqrt{b_1+b_2}}{2(\lambda+a)} \,,\qquad\displaystyle \hat{L}_{12}= \frac{e}{4}\,,\nn \\
\nn\\
\hat{L}_{21}&=&\scriptstyle\frac{(ab_1-b_2\lambda)\sin\phi+ab_2-b_1
\lambda}{\sin\theta\cos^2\phi(\lambda^2-a^2)}+ \frac{(a\sin\phi-\lambda)p_\phi^2}{\sin\theta(\lambda^2-a^2)}
 -\frac{2a\cos\phi\, p_\theta p_\phi}{\cos\theta(\lambda^2-a^2)}-\frac{\tan\theta(a\sin\phi+\lambda)p_\theta^2}{\cos\theta(\lambda^2-a^2)}-1\,,\nn\\
 \nn\\
 \hat{L}_{22}&=&\scriptstyle -\frac{a\cos\phi\,p_\phi}{\lambda^2-a^2}+\frac{\tan\theta(\lambda+a\sin\phi)p_\theta}{\lambda^2-a^2}-\frac{\mathrm i \sqrt{b_1-b_1}}{2(\lambda-a)}-\frac{\mathrm i\sqrt{b_1+b_2}}{2(\lambda+a)}\nn
 \en
 As above, zeroes of the both off-diagonal elements of $\hat{L}(\lambda)$ are the coordinates of separation for the Ha\-mil\-ton-Jacobi equation associated with the Hamilton function
\[
H_1=\dfrac{p_\phi^2}{\sin\theta}+\dfrac{\tan\theta(4\sin\theta+1)p_\theta^2}{\cos\theta}-4a\sin\phi\sin\theta+
\frac{b_1}{\sin\theta\cos^2\phi}+\frac{b_2\sin\phi}{\sin\theta\cos^2\phi}
\]
which is a coefficient in the equation defining the spectral curve
\bq\label{c-kow}
C:\quad 4\lambda(\lambda^2-a^2)\mu^2+\lambda^3-H_1\lambda+H_2=0\,.
\eq
If we substitute zeroes $v_{1,2}$ of the entry $\hat{L}_{21}(\lambda)$ and the corresponding momenta \[p_{v_{1,2}}=\hat{L}_{11}(v_{1,2})\] into the  two copies of the separated relation
\bq\label{c-kow2}
\begin{array}{l}
\tilde{C}:\quad\left(2(\lambda^2-a^2)\mu^2+2\mathrm i \bigl(\sqrt{b_1+b_2}(\lambda-a)+\sqrt{b_1-b_2}(\lambda+a)\bigr)\mu+\tilde{H}_1+\sqrt{\tilde{H}_2}+b_1^2-b_2^2\right)\qquad\\
\\
\qquad\times\left(2(\lambda^2-a^2)\mu^2+2\mathrm i \bigl(\sqrt{b_1+b_2}(\lambda-a)+\sqrt{b_1-b_2}(\lambda+a)\bigr)\mu+\tilde{H}_1-\sqrt{\tilde{H}_2}+b_1^2-b_2^2\right)\\
\\
\qquad\,\,-4\lambda^2+4b_3\lambda-8\sqrt{b_4}(\lambda^2-a^2)\mu=0\,.
\end{array}
\eq
and solve the resulting equations we get the Hamilton function for the generalized Kowalevski top
\bq\label{ham-kow}
\tilde{H}_1=J_1^2+J_2^2+2J_3^2+2ax_1-\dfrac{b_4}{x_3^2}+\dfrac{b_3}{\sqrt{x_1^2+x_2^2}}
+\dfrac{2-x_3^2}{x_2^2}\left(b_1+\dfrac{b_2x_1}{\sqrt{x_1^2+x_2^2}}\right)
\eq
after the following canonical transformation
\[
J_1=J_1+\dfrac{\sqrt{b_4\,}x_2}{x_3\sqrt{x_1^2+x_2^2}}\,,\qquad J_2=J_2-\dfrac{\sqrt{b_4\,}x_1}{x_3\sqrt{x_1^2+x_2^2}}\,.
\]
As above the separated relation (\ref{c-kow2}) defines genus 3 non-hyperelliptic curve.

These $v$-variables and  separated relation (\ref{c-kow2}) were obtained in \cite{ts13a,ts10k} in the framework of  bi-Hamiltonian geometry. Substituting $\hat{L}(\lambda)$ into the linear $r$-matrix equation (\ref{rp}) with $r$-matrices (\ref{r-3}) or (\ref{r-4}) we can recover the corresponding Poisson bivectors $P'_1$ and $P'_2$ compatible with the canonical bivector $P$ in $T^*\mathbb S^2$.

\section{Conclusion}
In this note we present a special set of variables on $T^*\mathbb S^2$ which are the variables of separation for the Neumann and Chaplygin systems simultaneously. These Chaplygin  variables can be obtained from the standard elliptic coordinates on the sphere using  the  B\"{a}cklund transformation.

Then we prove that similar  B\"{a}cklund transformations for other curvilinear coordinates on $T^*\mathbb S^{2}$ and $T^*\mathbb R^{2}$ yield variables of separation for the system with quartic potential, for the H\'{e}non-Heiles system and for the Kowalevski top. The corresponding separated relations  are associated with the genus-3 non-hyperelliptic curves, which bear a great resemblance to each other.

Thus, we start with  a lot of isolated known examples of two-dimensional separable systems and  prove that variables of separation for different systems  may be related by the B\"{a}cklund transformation. It allows us to say about some analog of the hetero B\"{a}cklund transformations relating different Hamilton-Jacobi equations.  Of course, on the next step we have to  explain the phenomenology described in this paper.

This work was partially supported by RFBR grant 13-01-00061.


\begin{thebibliography}{10}
\bibitem{harn88}
M. R. Adams, J. Harnad, and E. Previato,
\newblock{\em Isospectal Hamiltonian Flows in Finite and Infinite Dimensions}, Comm. Math. Phys., v.117, 451-500, 1988.

\bibitem{harn93}
M. R. Adams, J. Harnad and J. Hurtubise,
\newblock{\em Darboux Coordinates and Liouville-
Arnold Integration in Loop Algebras}, Commun. Math. Phys., v. 155, p. 385-
413, 1993.



\bibitem{bab92}
O. Babelon and M. Talon,
\newblock{\em Separation of variables for the classical and quantum Neumann model},
Nucl. Phys.B., v.379, p.321--339, 1992.


\bibitem{ped04}
 C. Bartocci, G. Falqui and M. Pedroni,
 \newblock{\em A geometric approach to the separability of the Neumann-Rosochatius system},
 Diff. Geom. Appl., v. 21, p. 349-360, 2004.

\bibitem{bell05}
M. P. Bellon and M. Talon,
\newblock{\em Spectrum of the quantum Neumann model}, 	Phys.Lett. A, v.337, p.360-368, 2005.




\bibitem{bob09}
 A.I. Bobenko, Yu.B. Suris. Discrete differential geometry: integrable structure. AMS, Providence, 2009. 8, 47, 230.

 \bibitem{bps13}
R. Boll, M.  Petrera,  and  Y.B. Suris, {Multi-time Lagrangian 1-forms for families of B\"{a}cklund transformations: Toda-type systems}, J. Phys. A: Math. and Theor., v.46, n.27, 275204, 2013.

\bibitem{ch03}
S. A. Chaplygin,
 \newblock{\em A new partial solution of the problem of motion of a rigid body in a liquid},
 Trudy otdel. Fiz. Nauk Obsh. Liub. Est., v.11, p. 7-10, 1903.


\bibitem{ves01}
H. R. Dullin, P. H. Richter, A. P. Veselov and H. Waalkens,
\newblock{\em Actions of the
Neumann systems via Picard-Fuchs equations}, Physica D, v.155, 159-183, 2001.




\bibitem{fordy91}
A. P. Fordy,
\newblock{\em The H\'{e}non-Heiles system revisited},
 Phys. D, v. 52, no. 2-3, p.204-210., 1991.

\bibitem{ts11a}
Yu. A. Grigoryev and A.V. Tsiganov,
\newblock{\em Separation of variables for the generalized H\'{e}non-Heiles system and system with quartic potential},
J. Phys. A: Math. Theor., v.44, 255202, 2011.


\bibitem{ts13a}
Yu. A. Grigoryev, V. A. Khudobakhshov and A.V. Tsiganov,
\newblock{\em	Separation of variables for some systems with a fourth order integral of motion},
Theor. Math. Phys., v.177(3), p. 1678-1690, 2013.

\bibitem{gs84}
V. Guillemin and S. Sternberg,
\newblock{\em Symplectic techniques in physics}, Cambridge, 1984.

\bibitem{gur95}
D. Gurarie,
\newblock{\em Quantized Neumann problem, separable potentials on $S^n$ and the Lam\'{e} equation},
J. Math.l Phys., v. 36, p.5355-5391, 1995.

\bibitem{jac66}
C. G. J. Jacobi, Vorlesungen \"{u}ber Dynamik, Georg Reimer, Berlin, 1866. Jacobi's
lectures on dynamics given in K\"{o}nigsberg 1842-1843 published by A. Clebsch.



\bibitem{ts11r}
V.A. Khudobakhshov and A.V. Tsiganov, \newblock{\em Integrable systems on the sphere associated with genus three algebraic curves}, Reg. Chaot. Dyn., v. 16 (3-4), p. 396 - 414, 2011.





\bibitem{kuz02s}
V.~B. Kuznetsov,
\newblock {\em Simultaneous separation for the Kowalevski and Goryachev–Chaplygin gyrostats },
J. Phys. A: Math. Gen., v.35, p.6419-6430, 2002.

\bibitem{kuz02}
V.~B. Kuznetsov, P. Vanhaecke,
\newblock{\em B\"{a}cklund transformations for finite-dimensional integrable systems: a geometric approach}, J. Geom. Phys. v.44, p.1-40, 2002.



\bibitem{lih77}
 A. Lichnerowicz,
\newblock{\em
Les varietes de Poisson et leurs algebres de Lie associees}, J. Diff. Geom., 1977, v.12, p.253-300.

\bibitem{li14}
Y. C. Li, A. Yurov,
Lie-B\"{a}cklund-Darboux Transformations, Surveys of Modern Mathematics, v.8, International Press, Somerville, MA; Higher Education Press,  Beijing, 2014.

\bibitem{mag97}
F. Magri,
\newblock{\em Eight lectures on Integrable Systems},
 {Lecture Notes in Physics, Springer Verlag, Berlin-Heidelberg}, v.495, p.256-296, 1997.




 \bibitem{mum84}
 D. Mumford,
 \newblock{Tata Lectures on Theta II}, Birkh\"{a}user, Boston, 1984.

 \bibitem{mos80-1}
 J. Moser, Geometry of quadrics and spectral theory, The Chern Symposium 1979 (Proc. Internat.
Sympos., Berkeley, Calif., 1979), Springer, New York, 1980, pp. 147-188.

\bibitem{mos80-2}
J. Moser,
\newblock{\em Various aspects of integrable Hamiltonian systems}. In: Dynamical systems (C.I.M.E. Summer School, Bressanone, 1978), Progr. Math. 8, Birkh\"{a}user, Boston, Mass., 1980, pp. 233-289.

\bibitem{neu59}
C. Neumann,
\newblock{\em De problemate quodam mechanico, quod ad primam integralium
ultraellipticorum classem revocatur}. Jour. Reine Angew. Math., v. 56, p. 46-63, 1859.


\bibitem{rav93}
V. Ravoson, L. Gavrilov and R. Caboz,
\newblock{\em Separability and Lax pairs for H\'{e}non-Heiles
system}, J. Math. Phys., v.34, no. 6, p.2385-2393, 1993.

\bibitem{rod02}
C. Rogers and W.K. Schief, B\"{a}cklund and Darboux transformations: geometry and modern
applications in soliton theory, vol. 30, Cambridge University Press, 2002.

\bibitem{rom95}
F. J. Romeiras,
\newblock{\em Separability and Lax pairs for the two-dimensional Hamiltonian system with a quartic potential },
J. Math. Phys. v.36, p.3559, 1995.


\bibitem{ros77}
E. Rosochatius,
\newblock{\"{U}ber die Bewegung eines Punktes}, Inaugural Dissertation, Univ. G\"{o}ttingen, Berlin, 1877.


\bibitem{skl95}
E.~K. Sklyanin
\newblock {\em Separation of variables---new trends}.
\newblock Progr. Theoret. Phys. Suppl., v.118, p. 35--60, 1995.




\bibitem{tt12}
P. Tempesta, G. Tondo,
 \newblock{\em Generalized Lenard chains, separation of variables, and superintegrability}, Phys. Rev. E, v.85, 046602, 11 pages, 2012.



\bibitem{ts04sl} A.V. Tsiganov,
\newblock{\em	On the Steklov-Lyapunov case of the rigid body motion},
Regular and Chaotic Dynamics, v.9(2), p.77-91, 2004.

\bibitem{ts04}
A.V. Tsiganov,
\newblock{\em 	Toda chains in the Jacobi method},
Teor. Math. Phys., v.139(1), p.636-652, 2004.

\bibitem{ts10a}
A.V. Tsiganov,
\newblock{\em On the generalized Chaplygin system},
Journal of Mathematical Sciences, v.168, n.8, p.901-911, 2010.

\bibitem{ts10k}
A.V. Tsiganov, \newblock{\em New variables of separation for particular case of the Kowalevski top},
Regular and Chaotic Dynamics, v.15, n.6, p. 657-667, 2010.

\bibitem{ts11}
A.V. Tsiganov, \newblock{\em On bi-integrable natural hamiltonian systems on Riemannian manifolds},
Journal of Nonlinear Mathematical Physics, v.18, n.2, p. 245-268, 2011.

\bibitem{ts11s} A.~V. Tsiganov,
\newblock{\em On natural Poisson bivectors on the sphere}, 	
J. Phys. A: Math. Theor., 44, 105203 (15pp), 2011.

\bibitem{ts12} A.~V. Tsiganov,
\newblock{\em	New variables of separation for the Steklov-Lyapunov system},
Symmetry, Integrability and Geometry: Methods and Applications , v.8, 012, 14 pages, 2012.


\bibitem{tur92}
F. Turiel,
{\em Structures bihamiltoniennes sur le fibr\'e cotangent},
C.\ R.\ Acad.\ Sci.\ Paris S\'er.\ I Math., v. 315, p. 1085--1088, 1992.

\bibitem{stef82}
S. Wojciechowski,\newblock{\em  The analogue of the B\"{a}cklund transformation for inte-
grable many-body systems}, J. Phys. A: Math. Gen. , v.15,  pp.653-657, 1982.

\bibitem{stef85}
S. Wojciechowski,
\newblock{\em Integrable one-particle potentials related to the Neumann system
and the Jacobi problem of geodesic motion on an ellipsoid}, Phys. Lett. A, v.107, p.106-111, 1985.


\bibitem{zul13}
F. Zullo, \newblock{\em  B\"{a}cklund transformations and Hamiltonian flows},
J. Phys. A: Math. and Theor., v.46, n.14, 145203, 2013.


\end{thebibliography}
\end{document}